# Proposal for a computation procedure based on continuous-variable post-selected teleportation.

Roberto Bovolenta*

*Piazza Matteotti 394, 45038, Polesella (RO), Italy.*

**Abstract:** in this paper, we analyze a potential procedure based on the combination of an alternative information encoding system inside the state of an electromagnetic mode and on continuous-variable post-selected teleportation, for speed up computational power of a classic architecture up to resolution of problems in PSPACE.

## I. INTRODUCTION.

In the environment of quantum computation, one of the milestone is certainly the quantum teleportation process developed by Bennett *et al.* in 1993 [1], which allow, under some restrictions, to transfer a state in an arbitrary distant point through entanglement sharing between two users, and further unitary operation by the side of the receiving (called "Bob").

As pointed out later by Lloyd *et al.* [2] [3], in the specific case where Bob doesn't need to perform any unitary operation on his half of the entangled state, we'll have a post-selection condition, where the half of Bob is projected into the initial state of the system even before this state is available, as showed in fig. 1. This peculiarity is called "closed timelike curve via quantum post-selection", or P-CTC, and it take inspiration from theory of Closed Timelike Curves (CTC) admitted by Einstein's theory of general relativity [4]. As pointed out by Aaronson e Watrous [5], CTCs allows the solution of any problem in PSPACE, whereas P-CTCs combined with quantum algorithms allows solution in PP, that is problems that a probabilistic polynomial Turing machine accepts with probability ½ if and only if the answer is "yes".

In this paper, a potential procedure will be analyzed based on the combination of an alternative information encoding system inside the state of an electromagnetic mode and on continuous-variable post-selected teleportation; this procedure, applied on a classic computer, it would allows the resolution of problems in PSPACE.

*E-mail: roberto.bovolenta@hotmail.it
        bovolenta.roberto@gmail.com

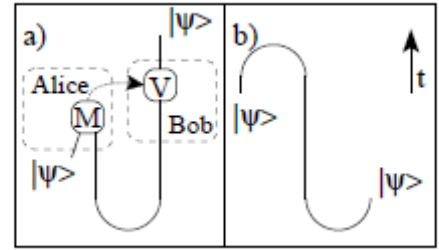

Fig. 1 – Description of closed timelike curves through teleportation. a) Conventional teleportation: Alice and Bob start from a maximally entangled state shared among them represented by "S". Alice performs a Bell measurement M on her half of the shared state and on the unknown state she wants to transmit. This measurement tells her which entangled state the two systems are in. She then communicates (dotted line) the measurement result to Bob who performs a unitary V on his half of the entangled state, obtaining the initial unknown state. b) Post-selected teleportation: the system in state and half of the Bell state "S" are projected onto the same Bell state "T". This means that the other half of the Bell state is projected into the initial state of the system even before this state is available.

## II. CONTINUOUS-VARIABLE POST-SELECTED TELEPORTATION.

Continuous-variable (CV) teleportation is achieved by means of exploitation of entanglement produced by combination of two squeezed states in a half beam splitter [6]. Since maximal entanglement between two squeezed states isn't a physical state, because it would require an unlimited squeezing level, we'll have necessarily a flaw in the teleportation process, estimated by fidelity
$F = <\psi_{in}|\rho_{out}|\psi>$, with $|\psi>$ input state and $\rho_{out}$ density operator for the teleported state.
In the following, we describe the teleportation process [7] in the Heisenberg representation. Initially, the sender Alice and the receiver Bob share a pair of EPR beams.

Alice performs a joint measurement on her EPR mode ($x_A, p_A$), and the input mode ($x_{in}, p_{in}$). She combines these two modes at a half beam splitter and measure, with two homodyne detection:

$$x_u = (x_{in} - x_A)/\sqrt{2}, \quad p_v = (p_{in} + p_A)/\sqrt{2}. \quad (1)$$

The measurement results ($x_u, p_v$) are then sent to Bob through classical channels with gain $g_x$ and $g_p$. Bob receives Alice's measurement results ($x_u, p_v$) through the classical channels and displaces his EPR beam ($x_B, p_B$) accordingly:

$$x_B \rightarrow x_{out} = x_B + \sqrt{2} x_u,$$
$$p_B \rightarrow p_{out} = p_B + \sqrt{2} p_v. \quad (2)$$

So, the teleported mode can be written as [8]:

$$x_{out} = x_{in} - (x_A - x_B),$$
$$p_{out} = p_{in} + (p_A + p_B). \quad (3)$$

Ideally, the EPR beams would have perfect correlations such that $x_A - x_B \rightarrow 0$ and $p_A + p_B \rightarrow 0$.

Hence, the teleported output would be identical to the input. In a real experimental situation, EPR beams have finite correlation and the variance would be written as $<[\Delta(x_A - x_B)]^2> = <[\Delta(p_A + p_B)]^2> = 2e^{-2r}\sigma_{vac}$. Here $\sigma_{vac}$ is a variance of vacuum fluctuation, and $r$ is a squeezing parameter.

So, to achieve CV post-selected teleportation, it's necessary that $x_u = 0$, namely $x_{in} = x_A$. A feasible realization scheme is showed in fig. 2 and 3. First of all, we produce two couples of identical squeezed states [9], by means of a parametric amplifier and beam splitters (fig. 2).

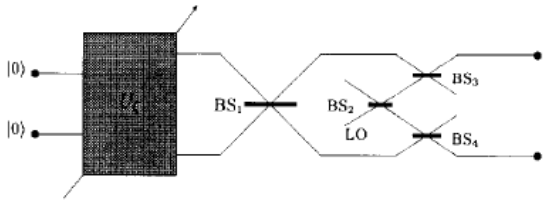

Fig. 2 – Schematic diagram of the interaction scheme for generating two identical squeezed states. The grey box represents a parametric amplifier along with its classical pump. $BS_1$ and $BS_2$ are balanced beam splitters whereas $BS_3$ and $BS_4$ are nearly-unit transmissivity beam splitters. LO denotes a local oscillator, namely a very intense laser beam.

Thereafter, we combine them in two other half beam splitters: a component of a couple of squeezed states combines with a component of the other couple, so that the input state and the EPR half of Alice remain the same. Finally, Alice carry out the homodyne detection among the input state and her EPR half (fig. 3). At this stage, there will be no need to implement the displacement operation on the output state, which will result the same – except for not perfect fidelity – to the input state, even *before* the homodyne detection.

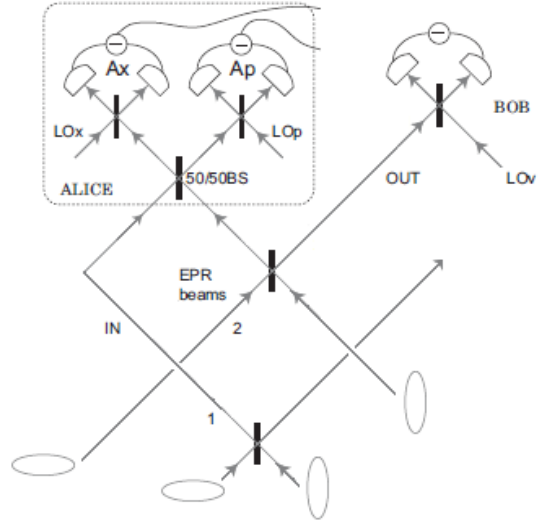

Fig. 3 – Schematic setup for post-selected CV teleportation. First we produce two couples of identical squeezed states, then we combine them by means of two half beam splitters: a component of a couple of squeezed states combines with a component of the other couple, so that the input state and the EPR half of Alice remain the same. Finally, Alice carry out the homodyne detection among the input state and her EPR half, and there will be no need to implement the displacement operation on the output state, which will result the same – except for not perfect fidelity – to the input state, even *before* the homodyne detection.

### III. INFORMATION ENCODING.

In CV quantum computation, the electromagnetic modes, called qumodes, can be exploited for encoding qubits [10,11], or in a direct computation [12,13]. Now we'll show an alternative information encoding system inside qumodes, which will make use of a

single logic gate:
the Pauli operator $\mathbf{X}(x) \equiv e^{-2ix\mathbf{p}}$, which act on position operator as [13]:

$$\mathbf{X}(x')|x\rangle = |x + x'\rangle. \quad (4)$$

Now, let $\mathbf{x}_0$ be the initial state of our qumode, we set a point L and divide the segment $L - \mathbf{x}_0$ in n sections, each of length $\alpha$, so $L - \mathbf{x}_0 = n\alpha$ (fig. 4). Now, for each section, we link a specific input, which can be either a string of bits or a m × n matrix which represent a specific memory configuration in terms of cells and columns. The choice of L and $\alpha$ will depend on the computation input dimension and/or on the available memory of "classic" instrumentation.

So, for example, if the initial configuration corresponding to input data is located in interval $\alpha_5$ (we define $\alpha_0$ first interval), we'll perform the following operation:

$$\mathbf{X}(s)|x_0\rangle = |x_0 + s\rangle, \quad 5\alpha < s < 6\alpha. \quad (5)$$

Alternatively, we can add encoding for $\mathbf{p}$ too, therefore employing the $\mathbf{Z}(p) \equiv e^{2ip\mathbf{x}}$ operator, which has the following property:

$$\mathbf{Z}(p)|x\rangle = e^{2ip\mathbf{x}}|x\rangle. \quad (6)$$

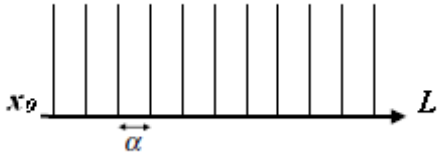

Fig. 4 – Segment $[x_0, L]$ and intervals $\alpha_n$ corresponding to different inputs. Example: $\alpha_0$ could corresponding to 0, $\alpha_{12}$ to binary string 1100, or to $\begin{pmatrix} 100 \\ 001 \\ 000 \end{pmatrix}$ matrix, etc.

By convention, we can choose $\mathbf{X}$ in order to $\mathbf{x}$ fall in the midpoint of each interval, so that we'll have a tolerance range of $\pm 0.5\alpha$.

## IV. IMPLEMENTATION OF CV P-CTC TO COMPUTATION.

The final scheme is showed in fig. 5. First of all, we implement the CV post-selected teleportation scheme; then, we "load" data on our input state by means of Pauli operator (or operators).

So, we call $\mathbf{x}_i$ the chosen translation value corresponding to data input. Such operation will be implemented also to Alice's EPR half, since the two states must have same $\mathbf{x}$. At an instant t previous such operation, we'll make the tomographic measure of Wigner function of Bob state [14], from which we'll obtain the value of $\mathbf{x} = \mathbf{x}_0 + \mathbf{x}_i$ (and, possibly, of $\mathbf{p}$).

Inside a classic computer, it will make the comparison between the value of $\mathbf{x}$ and the input to compute, in order to carry out $\mathbf{x}_0$, L and the intervals $\alpha_n$.

Now, we call $q_0$ the initial state of computation in which are loaded the inputs and $q_1$ the state after the first computation cycle (or the clock cycle); then, $q_1$ state will be converted in the analogue $\alpha_k$ from $\mathbf{X}$ operator *before* the input state pass through the above-mentioned logic gate.

When the computation will reach the final state $q_f$, this will not be changed anymore in the time loop and it will be carried outwards and read as output.

Now, although every $\alpha_n$ include an unlimited number of values, for a matter of convention and pragmatism, the value of $\mathbf{X}$ could be chosen so that $\mathbf{x}$ fall on the half of each interval.

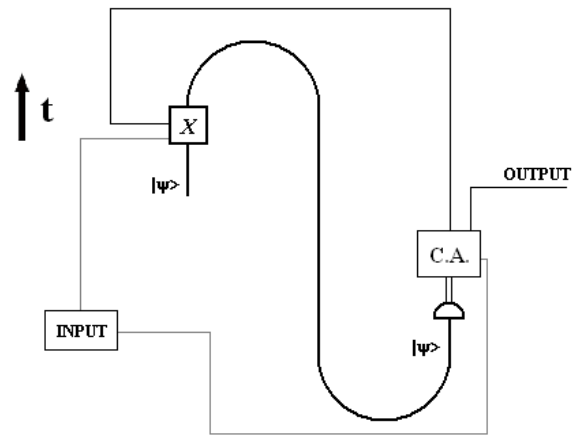

Fig. 5 – First scheme for implementation of CV P-CTC to a computational process. Inputs – said "$q_0$ state" – are loaded both in the electromagnetic mode and in the classic computer, so that we can obtain $\mathbf{x}_0$, L and the intervals $\alpha_n$. After the first computation cycle (or clock cycle), $q_1$ state will be converted in the analogue

operation **X** *before* the entrance of input state, in order to create the time loop which instantly will deliver to the final state $q_f$. The $q_f$ state will not change anymore, and will be read in output.

## V. ALTERNATIVE CIRCUIT.

Now, we show in fig. 6 an alternative scheme of our procedure. First, we implement again our CV post-selected teleportation scheme, then we produce another couple of identical squeezed states. Each of the two states will be joined, respectively, with the input and output state by means of a QND (Quantum Non-Demolition) gate [7]. The QND gate makes interaction between two input modes with a Hamiltonian $\mathbf{H}_{QND} = \mathbf{x}_1\mathbf{p}_2$. Input and output relation is obtained as:

$$\mathbf{x}_1^{out} = \mathbf{x}_1^{in}$$
$$\mathbf{x}_2^{out} = \mathbf{x}_2^{in} + G\mathbf{x}_1^{in}$$
$$\mathbf{p}_1^{out} = \mathbf{p}_1^{in} - G\mathbf{p}_2^{in}$$
$$\mathbf{p}_2^{out} = \mathbf{p}_2^{in}. \qquad (7)$$

The procedure works in the following: we call $\mathbf{x}_{in}$ and $\mathbf{x}_{out}$ the position operators of input and output states, and $\mathbf{x}_{sq}$ that of our new squeezed states, which function as ancilla states (so, after QND gate, $\mathbf{x}_{in}^{out} = \mathbf{x}_{in}^{in}$ and $\mathbf{x}_{out}^{out} = \mathbf{x}_{out}^{in}$).

In a classic computer, it will be made the entire process of computation $q_0 \rightarrow q_f$, and at the same time it will be measured the value:

$$\mathbf{x}_{sq}^{out1} = \mathbf{x}_{sq}^{in} + G\mathbf{x}_{in}^{in}. \qquad (8)$$

When the computation process is complete, the $q_f$ state will be implemented in the input state by means of **X.** At an instant t previous such operation, it will be measured, at output, the value:

$$\mathbf{x}_{sq}^{out2} = \mathbf{x}_{sq}^{in} + G\mathbf{x}_{out}^{in}. \qquad (9)$$

Having obtained the value $\mathbf{x}_{sq}^{out1}$ before the computation process, we can extract the difference:

$$\Delta\mathbf{x}_{sq} = \mathbf{x}_{sq}^{out1} - \mathbf{x}_{sq}^{out2} = G(\mathbf{x}_{in}^{in} - \mathbf{x}_{out}^{in}). \qquad (10)$$

Therefore, the length $\mathbf{x}_{in}^{in} - \mathbf{x}_{out}^{in}$ allow us to locate the interval $\alpha_k$ corresponding to $q_f$.

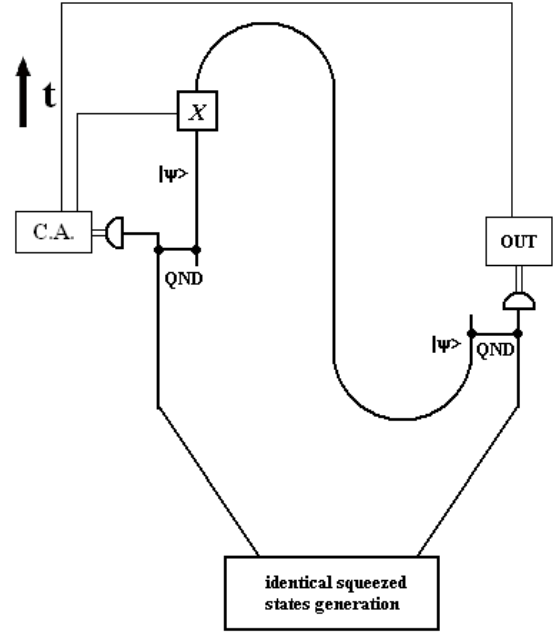

Fig. 6 – Second scheme for implementation of CV P-CTC to a computational process. In this case, the classic computer will make the entire process of computation $q_0 \rightarrow q_f$, and it will load the solution $q_f$ in the electromagnetic mode by means of **X**. The solution will be read before the end of computation, and two identical squeezed states will serve as ancilla states to obtain the length $\mathbf{x}_{in}^{in} - \mathbf{x}_{out}^{in}$ which allow us to locate the interval $\alpha_k$ corresponding to $q_f$.

## VI. CONCLUSIONS.

In this paper, we have showed a new potential procedure for implementation of CV P-CTC to computational processes. The peculiarity of this procedure is that, unlike other suggestions [2,3], here the quantum computation is restricted by only two logic gates, the Pauli operator **X** and the QND gate. The proper algorithm process is addressed to a classic computer which interacts with P-CTC by means of **X** and quantum tomographic measure.

We have showed two different schemes: the first refers to a feedback loop scheme [5] which uses, as quantum logic gate, only the **X** operator, and potentially susceptible to experimental errors which could compromise the entire process [2,3]. The second scheme, on the other hand, first performs the entire computational process in the classic computer, then the final state $q_f$ (solution of the problem) will be implemented in the input state by means of **X**. The output measure will be revealed in a *previous* instant at the end of computation.

The scheme uses also two QND gates to obtain, by means of two identical ancilla squeezed states, the length $\mathbf{x}_{in}^{in} - \mathbf{x}_{out}^{in}$ which allows us to obtain the interval $\alpha_k$ corresponding to $q_f$.

**ACKNOWLEDGEMENTS**
I gratefully acknowledge helpful discussions with Dr. Elio Fabri, Dr. Lorenzo Maccone and Dr. Vittorio Giovannetti.